\documentclass{article}

\usepackage{microtype}
\usepackage{graphicx}
\usepackage{subfigure}
\usepackage{booktabs} 

\usepackage{url}

\usepackage{hyperref}

\usepackage[accepted]{icml2023-diffxyz}

\usepackage{amsmath}
\usepackage{amssymb}
\usepackage{mathtools}
\usepackage{amsthm}

\usepackage[capitalize,noabbrev]{cleveref}

\theoremstyle{plain}

\theoremstyle{definition}

\theoremstyle{remark}

\usepackage[textsize=tiny]{todonotes}

\usepackage{lipsum}
\usepackage{enumitem}
\usepackage{duckuments}
\usepackage[font=small,skip=5pt]{caption}
\usepackage[frozencache,cachedir=.]{minted}
\setminted[python]{baselinestretch=1.0,fontsize=\footnotesize, baselinestretch=1.0}  
\icmltitlerunning{JAX FDM:~A differentiable solver for inverse form-finding}

\begin{document}

\twocolumn[
\icmltitle{JAX FDM:~A differentiable solver for inverse form-finding}



\icmlsetsymbol{equal}{*}

\begin{icmlauthorlist}
\icmlauthor{Rafael Pastrana}{yyy,sch}
\icmlauthor{Deniz Oktay}{comp}
\icmlauthor{Ryan P. Adams}{comp}
\icmlauthor{Sigrid Adriaenssens}{sch}
\end{icmlauthorlist}

\icmlaffiliation{yyy}{\{School of Architecture, }
\icmlaffiliation{sch}{Department of Civil and Environmental Engineering, }
\icmlaffiliation{comp}{Department of Computer Science\}, Princeton University, USA}

\icmlcorrespondingauthor{Rafael Pastrana}{arpastrana@princeton.edu}

\icmlkeywords{Structural design, Structural optimization, Differentiable physics, Automatic differentiation, Force density method, Form finding, JAX}

\vskip 0.3in
]



\printAffiliationsAndNotice{}  

\begin{abstract}
We introduce JAX FDM, a differentiable solver to design mechanically efficient shapes for 3D structures conditioned on target architectural, fabrication and structural properties.
Examples of such structures are domes, cable nets and towers.
JAX FDM solves these inverse form-finding problems by combining the force density method, differentiable sparsity and gradient-based optimization.
Our solver can be paired with other libraries in the JAX ecosystem to facilitate the integration of form-finding simulations with neural networks.
We showcase the features of JAX FDM with two design examples.
JAX FDM is available as an open-source library at this URL:\;\href{https://github.com/arpastrana/jax\_fdm}{https://github.com/arpastrana/jax\_fdm}
\end{abstract}


\section{Introduction}\label{intro}

The force density method (FDM)~\cite{schek_force_1974} is a form-finding method that generates shapes in static equilibrium for meter-scale 3D structures, such as masonry vaults~\cite{panozzo_designing_2013}, cable nets~\cite{veenendaal_structural_2017} and tensegrity systems~\cite{zhang_adaptive_2006}.
A structure in static equilibrium carries loads like its self-weight or wind pressure only through internal tension and compression forces~\cite{adriaenssens_shell_2014}.
This axial-dominant mechanical behavior enables a form-found structure to span long distances with low material usage compared to a structure that is not form-found~\cite{schlaich_shell_2018,rippmann_armadillo_2016}.

\begin{figure}[t!]
    \centering
    \includegraphics[width=\columnwidth]{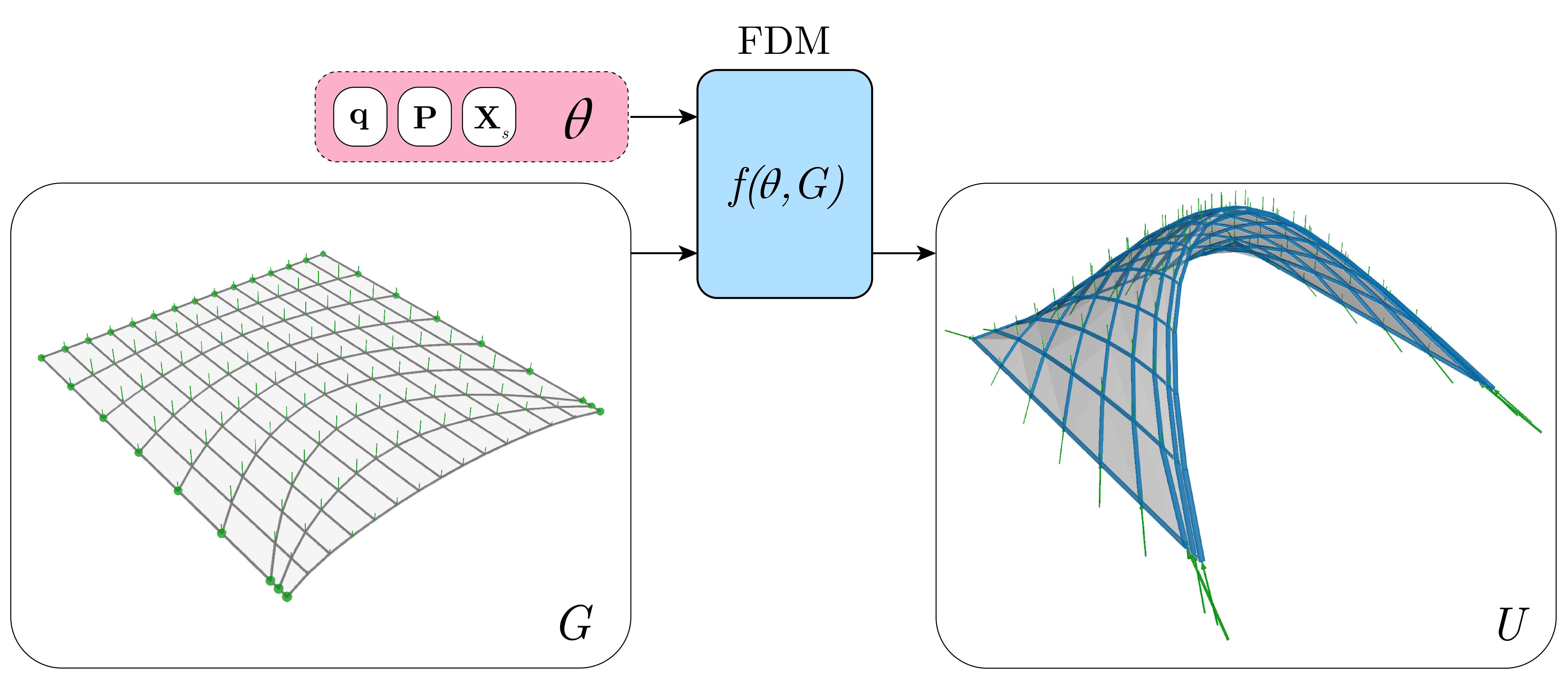}
    \caption{The FDM is a forward form-finding method that computes a state of static equilibrium $U$ on an input graph $G$ given input parameters $\theta=(\mathbf{q},\mathbf{P},\mathbf{X}_{\text{s}})$.}
    \label{fig:fdm_forward}
    \vskip -0.2in
\end{figure}

\begin{figure}[b!]
    \centering
    \includegraphics[width=\columnwidth]{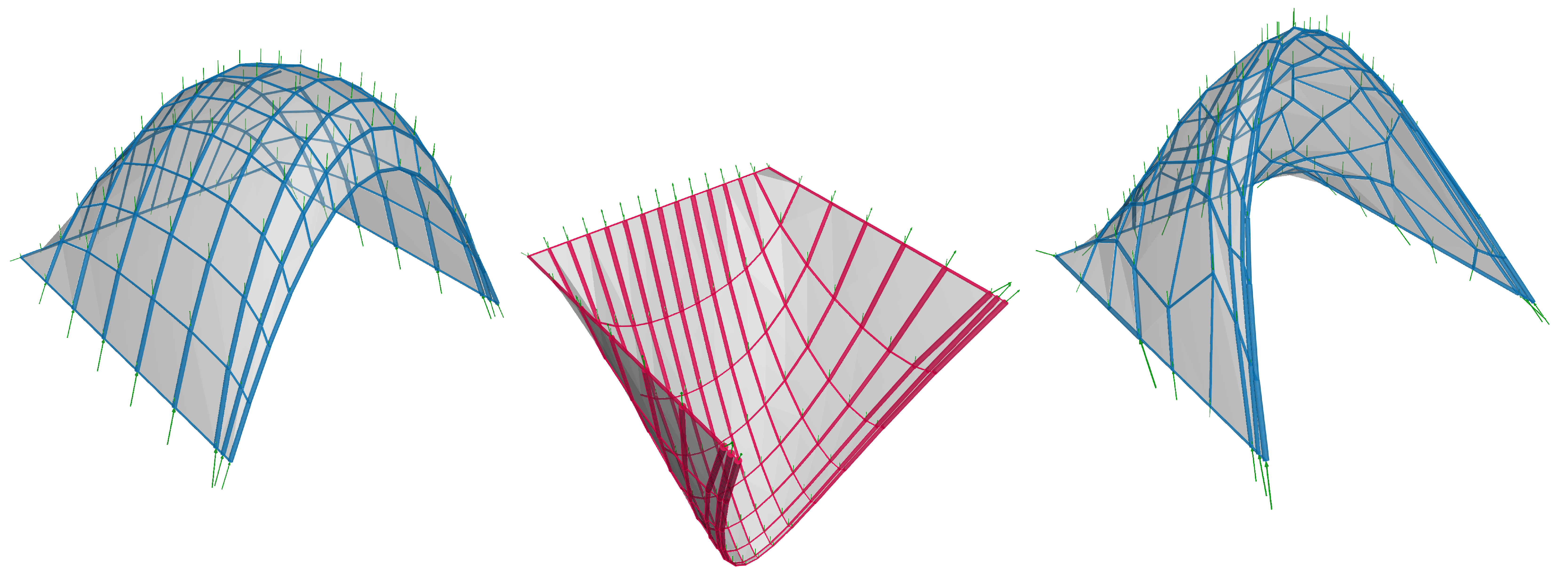}
    \caption{The FDM generates different static equilibrium configurations for variations of $\theta$. From left to right: $\mathbf{q}\in\{-0.1, -1\}$, $\mathbf{q}=\mathbf{1}$, and $\mathbf{q}\sim\mathcal{U}(-0.1, -1)$. Colors indicate the internal axial forces: blue denotes compression and red tension.}
    \label{fig:fdm_variations}
\end{figure}

The FDM is a \textit{forward} physics solver expressed as a function $f(\theta,G)=U$.
Given a structure modeled as a sparse graph $G$ and a set of continuous design parameters $\theta$, the FDM computes a state of static equilibrium $U$ for $G$~(Fig.~\ref{fig:fdm_forward}). 
By inputting different values of $\theta$, the FDM generates a variety of shapes in static equilibrium (Fig.~\ref{fig:fdm_variations}).
However, in engineering practice, it is necessary to generate not any mechanically efficient 3D shape, but a feasible one that satisfies constraints arising from architectural, fabrication, or other structural requirements.

\begin{figure}[t]
    \vskip 0.2in
    \centering
    \centerline{\includegraphics[width=0.55\columnwidth]{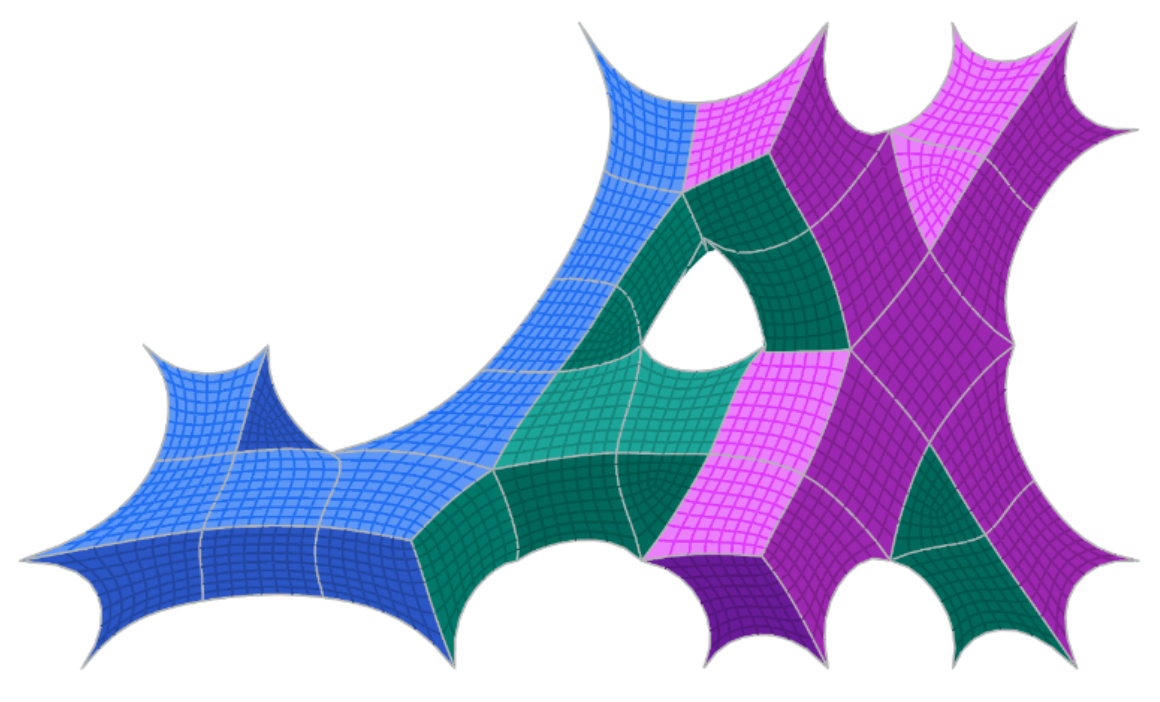}}
    \caption{Plan view of cable net with equalized edge lengths.}
    \label{fig:jax_logo}
    \vskip -0.2in
\end{figure}

Consider the case shown in Fig.~\ref{fig:fdm_inverse} where it is of interest to find a shape in equilibrium that is as close as possible to a target surface $\hat{U}$.
This surface may express architectural intent for a new roof or can represent the geometry of a historical masonry vault that needs to be analyzed for restoration purposes~\cite{panozzo_designing_2013,marmo_thrust_2019}.
Practical structural design, therefore, requires the solution of an \textit{inverse} form-finding problem, a mapping from~${\hat{U}\to\theta}$, where the goal is to estimate adequate values for the parameters $\theta^{\star}$ that are conducive to an equilibrium state with prescribed characteristics $\hat{U}$.

The design space of all possible shapes in static equilibrium parametrized by $\theta$ is vast, particularly as the dimensionality of these parameters grows proportionally to the hundreds or thousands of cables, bricks and blocks that compose a real-world structure.
Numerical approaches based on geometric heuristics~\cite{lee_automatic_2016} or genetic algorithms~\cite{koohestani_form-finding_2012} offer limited support to navigate this high-dimensional design space towards feasible designs.
The current surge of differentiable physics solvers and physics-informed neural networks in structural engineering~\cite{cuvilliers_constrained_2020,chang_learning_2020,xue_jaxfem_2023,wu_jaxsso_2023,pastrana_constrained_2023} provide insights to develop new approaches to tackle inverse form-finding.

In this paper, we present JAX FDM, a differentiable solver to perform inverse form-finding on 3D structures modeled as pin-jointed bar networks.
JAX FDM implements the FDM in JAX~\cite{bradbury_jax_2018} and solves inverse form-finding problems by estimating adequate inputs to the FDM via gradient-based optimization.
The required forward and backward calculations are executed efficiently by running a differentiable sparse solver on a CPU or a GPU.
After presenting the theory behind our work in Section~\ref{sec:method}, we use our solver to address two inverse form-finding problems: the design of a shell structure that matches an arbitrary target shape (Section~\ref{sec:results:shell}) and the design of cable net with prescribed edge lengths (Section~\ref{sec:results:cablenet}).
JAX FDM is open-source software accessible at this URL:\;\href{https://github.com/arpastrana/jax_fdm}{https://github.com/arpastrana/jax\_fdm}.

\section{Auto-differentiable and sparsified FDM}\label{sec:method}

\subsection{The force density method (FDM)}\label{sec:method:fdm}

The FDM models a structure as a pin-jointed, force network~\cite{schek_force_1974}.
Let $G=(V,E)$ be a graph with $n$ vertices $V$ connected by $m$ edges $E$ encoding this network.
One portion of $V$ of size $n_s$ is defined as the supported vertices of the structure $V_s$, i.e., the locations in the structure that are fixed and transfer reaction forces to its anchors.
The remaining $n_u$ unsupported vertices are denoted $V_u$.

A connectivity matrix $\mathbf{C}\in\{-1,0,1\}^{m\times n}$ encodes the relationship between the edges and the vertices of $G$.
Entry $c_{ij}$ of $\mathbf{C}$ is equal to $1$ if vertex $j$ is the start node of edge~$i$ and equal to $-1$ if vertex $j$ is the end node of edge $i$. Otherwise, $c_{ij}=0$.
Submatrices $\mathbf{C}_u$ and $\mathbf{C}_s$ are formed by the columns of $\mathbf{C}$ corresponding to the unsupported and the supported vertices of $G$, respectively.

The FDM is a function $f$ that computes a state of static equilibrium $U$ on a fixed graph $G$ given input parameters $\theta$.
The input parameters $\theta=(\mathbf{q},\mathbf{P},\mathbf{X}_{\text{s}})$ are features defined on the elements of $G$:
\begin{itemize}[topsep=1pt,itemsep=1pt,partopsep=1pt,parsep=1pt]
    \item A diagonal matrix $\mathbf{Q}\in\mathbb{R}^{m\times m}$ with the force densities of the edges, $\mathbf{q}\in\mathbb{R}^{m\times 1}$. The force density $q_{i}$ of edge $i$ is the ratio between the internal force $t_{i}$ and the length $l_{i}$ of the edge, $q_{i}=t_{i}/l_{i}$. A negative $q_i$ indicates compression while a positive one indicates tension.
    \item A matrix $\mathbf{P}\in\mathbb{R}^{n\times3}$ with the 3D vectors denoting the external loads applied to all the vertices of $G$. Submatrices $\mathbf{P}_u\in\mathbb{R}^{n_u\times3}$ and $\mathbf{P}_s\in\mathbb{R}^{n_s\times3}$ correspond to the rows of $\mathbf{P}$ with the loads applied to the vertices $V_u$ and $V_s$, respectively.
    \item A matrix $\mathbf{X}_s\in\mathbb{R}^{n_s\times3}$ containing the 3D coordinates of the supported vertices, $V_s$.
\end{itemize}

\begin{figure}[t!]
    \centering
    \vskip 0.38in
    \includegraphics[width=\columnwidth]{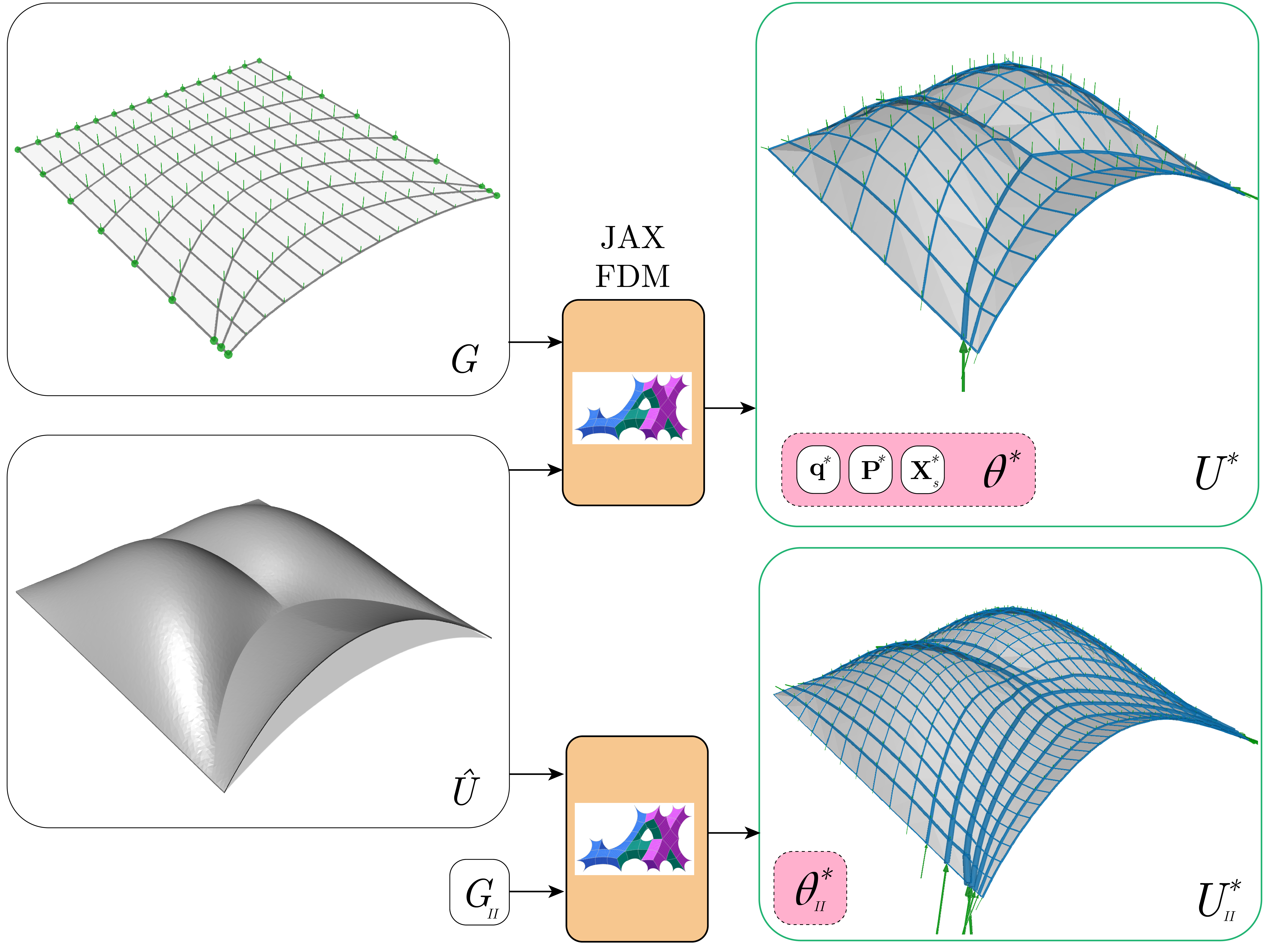}
    \caption{JAX FDM solves inverse form-finding problems estimating parameters $\theta^{\star}$ that fit a prescribed equilibrium state $\hat{U}$ via gradient-based optimization. Here, we calculate the force densities $\mathbf{q}^{\star}$ needed to solve a shape approximation problem on a creased shell modeled as graphs $G$ and $G_{II}$.}
    \vskip -0.2in
    \label{fig:fdm_inverse}
\end{figure}

The state $U=(\mathbf{X}_u,\mathbf{R}_s,\mathbf{t},\mathbf{l})$ characterizes the static equilibrium configuration of $G$:
\begin{itemize}[topsep=1pt,itemsep=1pt,partopsep=1pt,parsep=1pt]
    \item A matrix $\mathbf{X}_u\in\mathbb{R}^{n_u\times3}$ containing the 3D coordinates in static equilibrium of the unsupported vertices, $V_u$.
    \item A matrix $\mathbf{R}_s\in\mathbb{R}^{n_s\times3}$ with the reaction forces incident to the supported vertices, $V_s$.
    \item A vector $\mathbf{t}\in\mathbb{R}^{m\times 1}$ containing the tensile or compressive internal force of the edges.
    \item A vector $\mathbf{l}\in\mathbb{R}^{m\times 1}$ with the edge lengths.
\end{itemize}

The key step in the FDM is to find the 3D coordinates in static equilibrium of the free vertices $\mathbf{X}_u$ with Eq.~\ref{eq:xyz_free}:
\begin{equation}\label{eq:xyz_free}
    \mathbf{X}_u = (\mathbf{C}_u^{\text{T}}\mathbf{Q}\mathbf{C}_u)^{-1}
    (\mathbf{P}_u - \mathbf{C}_u^{\text{T}}\mathbf{Q}\mathbf{C}_s\mathbf{X}_s)\,
\end{equation}

The remaining components of $U$ are computed as:
\begin{align}
    \mathbf{R}_s &= \mathbf{P}_s - \mathbf{C}_s^{\text{T}}\mathbf{Q}\mathbf{C}\mathbf{X}
    \label{eq:xyz_reactions} \\
    \mathbf{t} &= \mathbf{q}^{\text{T}}\mathbf{L} \label{eq:forces}
\end{align}
The matrix of 3D coordinates $\mathbf{X}$ results from concatenating $\mathbf{X}_u$ and $\mathbf{X}_s$.
The diagonal matrix of edge lengths $\mathbf{L}$ can be calculated taking the row-wise L2 norm of the inner product of the connectivity matrix $\mathbf{C}$ and $\mathbf{X}$, $\mathbf{L} = \text{diag}(\lVert\mathbf{C}\,\mathbf{X}\rVert_2)$.

\subsection{Solving inverse form-finding problems}\label{sec:method:inverse}

The desideratum is to design structures in static equilibrium that attain additional architectural, fabrication, or other target properties.

The FDM parametrizes form-finding in terms of $\theta$, simplifying the computation of a state of static equilibrium $U$ to the solution of a linear system. 
However, the relationship between $\theta$ and $U$ is non-linear as linear perturbations in $\theta$ do not correspond to linear changes in $U$.
Moreover, the force densities $\mathbf{q}$ are not interpretable quantities: they express a ratio between the expected forces and lengths in the edges of a structure, but neither of them concretely.
Both issues complicate tackling inverse form-finding problems without an automated approach.

To address these challenges, we solve an unconstrained optimization problem w.r.t.~parameters $\theta$.
Let $g(f(\theta,G))$ be a non-linear goal function that computes a property of interest.
One inverse form-finding problem may contain $K$ different goal functions that are individually scaled by a weight factor $w_k$ and aggregated in a loss function $\mathcal{L}(\theta)$:
\begin{equation}\label{eq:loss}
    \mathcal{L}(\theta)=\sum_{k=1}^{K}w_k\,g_k(f(\theta,G)))
\end{equation} 
We minimize Eq.~\ref{eq:loss} by estimating optimal parameters $\theta^{\star}$ via gradient descent, iteratively updating $\theta$ in the negative direction of the gradient $\nabla_{\theta}\mathcal{L}$.
We conveniently estimate the required value of $\nabla_{\theta}\mathcal{L}$ with reverse-mode automatic differentiation~\cite{bradbury_jax_2018}.

\subsection{Differentiable sparse solver}\label{sec:method:sparse}

A bottleneck in our solver is the solution of the linear system in Eq.~\ref{eq:xyz_free}. 
Although for small problems we can materialize and invert the full dense coefficient matrix $\mathbf C_u^T \mathbf Q \mathbf C_u$, for larger problems we want to take advantage of the inherent sparsity of $\mathbf C_u$ to get computational speedup, especially as the linear solve is called many times during inverse design.

As sparse solvers have limited support on JAX at the time of writing, we use implicit differentiation to derive a custom differentiable sparse linear solver.
Given that the sparsity pattern only depends on $\mathbf{C}$, which is fixed from the beginning for a given graph $G$, we implement a differentiable map from force densities into the entries of the coefficient matrix in compressed sparse-row (CSR) format.
We then use the adjoint method to implement a custom gradient for \texttt{scipy.sparse.linalg.spsolve} on CPU, and \texttt{jax.experimental.sparse.linalg.spsolve} on GPU. 
On CPU, we use a \texttt{jax.pure\_callback} to ensure the sparse solve is compatible with JIT compilation.


\section{Examples}

\begin{figure}[t]
    \centering
    \includegraphics[width=\columnwidth]{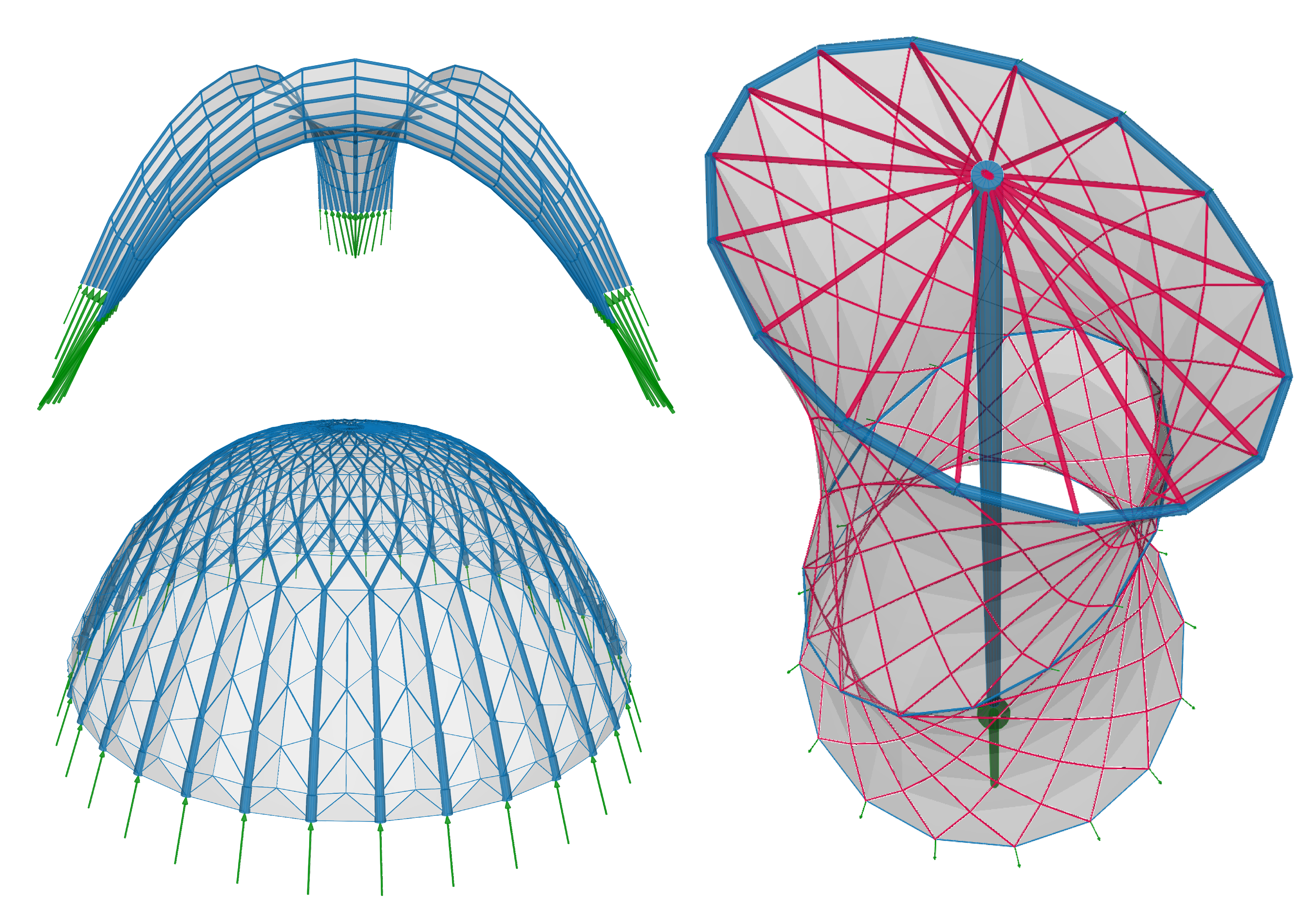}
    \caption{Inverse form-found structures generated with JAX FDM. Left: two compression-only structures, a monkey saddle (top) and a dome (bottom). Right: a tensegrity tower.}
    \label{fig:structures}
\end{figure}

JAX FDM features a rich bank of goal functions that simplify the modeling of inverse-form-finding problems on various structural systems (Fig.~\ref{fig:structures}).
Here, we present two specific use cases with the current version of the library.

\subsection{Shape approximation for shell structures}\label{sec:results:shell}

We want to calculate a network in static equilibrium for a shell that approximates the geometry $\hat{U}$ pictured in Fig.~\ref{fig:fdm_inverse}~\cite{panozzo_designing_2013}.
This target shape is supplied by the project architect as a COMPAS network~\cite{van_mele_compas_2017}.
JAX FDM offers functions to convert such a network into a JAX-friendly \texttt{jfe.EquilibriumStructure}.
This \texttt{structure} encodes the graph representation $G$ of the shell and its connectivity matrix $\mathbf{C}$ (Section~\ref{sec:method:fdm}).

\begin{minted}{python}
import jax_fdm.equilibrium as jfe
from jax_fdm.datastructures import FDNetwork

# load a structure from a COMPAS network 
net = FDNetwork.from_json("shell.json")
eqs = jfe.EquilibriumStructure
structure = eqs.from_network(net)
\end{minted}

To compute a state of static equilibrium \texttt{eq\_state} with the FDM, we instantiate an \texttt{jfe.EquilibriumModel} and define $\theta$ as a tuple of design parameters, \texttt{params}.

\begin{minted}{python}
import jax.numpy as jnp

# set the initial force densities
q = jnp.ones(structure.num_edges) * -1.0

# compute an equilibrium state
fdm = jfe.EquilibriumModel()
params = (q, xyz_fixed, loads)
eq_state = fdm(params, structure)
\end{minted}

The \texttt{fdm} model is a callable object that expresses $f(\theta,G)$ and implements Eqs.~\ref{eq:xyz_free}-\ref{eq:forces}.
The initial vector of force densities is set to $\mathbf{q}=-\mathbf{1}$.
The negative values denote compressive internal forces in the edges of $G$.
The other arrays, \texttt{xyz\_fixed} and \texttt{loads}, store the 3D coordinates of the supports $\mathbf{X}_s$, and the loads $\mathbf{P}$ applied to the vertices of $G$, respectively.
Next, we set up an inverse form-finding problem in terms of $\mathbf{q}$ with two functions:

\begin{minted}{python}
def goal_fn(eq_state):
    dist = (eq_state.xyz - xyz_target)**2
    return jnp.sum(dist) 

def loss_fn(q):
    params = (q, xyz_fixed, loads)
    eq_state = fdm(params, structure)
    return goal_fn(eq_state)
\end{minted}

The first one is a goal function $g(f(\theta,G))$, which quantifies the fitness of the shape approximation by measuring the cumulative distance between the \texttt{xyz} coordinates in static equilibrium of the vertices $V$ produced by \texttt{fdm}, and the \texttt{xyz\_target} coordinates on the objective surface.
The second function represents Eq.~\ref{eq:loss}, which we minimize with an \texttt{optax} optimizer~\cite{babuschkin_deepmind_2020}:

\begin{minted}{python}
import optax
from jax import jit
from jax import value_and_grad

@jit
def opt_step(q, o_state):
    loss, grad = value_and_grad(loss_fn)(q)
    upd, o_state = opt.update(grad, o_state)
    q = optax.apply_updates(q, upd)
    return q, o_state, loss

# optimization loop
opt = optax.adam(learning_rate=0.01)
o_state = opt.init(q)

for i in range(5000):
    q, o_state, loss = opt_step(q, o_state)
\end{minted}

The object abstractions and equilibrium calculations in JAX FDM are compatible with JAX transformations, such as \texttt{jit} and \texttt{value\_and\_grad}.
This compatibility allows us to write the optimization step for $\mathbf{q}$ with the same code blocks conventionally used to train neural networks.

Post-optimization, the distance between the solution provided by $\mathbf{q}^{\star}$ and the target shape decreases by four orders of magnitude.
The fit is comparable with an input graph $G_{II}$ that has three times more edges and design parameters (Fig.~\ref{fig:fdm_inverse}).
This example is available in a Colab notebook at \href{https://tinyurl.com/25czahvh}{https://tinyurl.com/25czahvh}.

\subsection{Equalizing edge lengths in a cable net}\label{sec:results:cablenet}

\begin{figure}[t]
    \centering
    \includegraphics[width=\columnwidth]{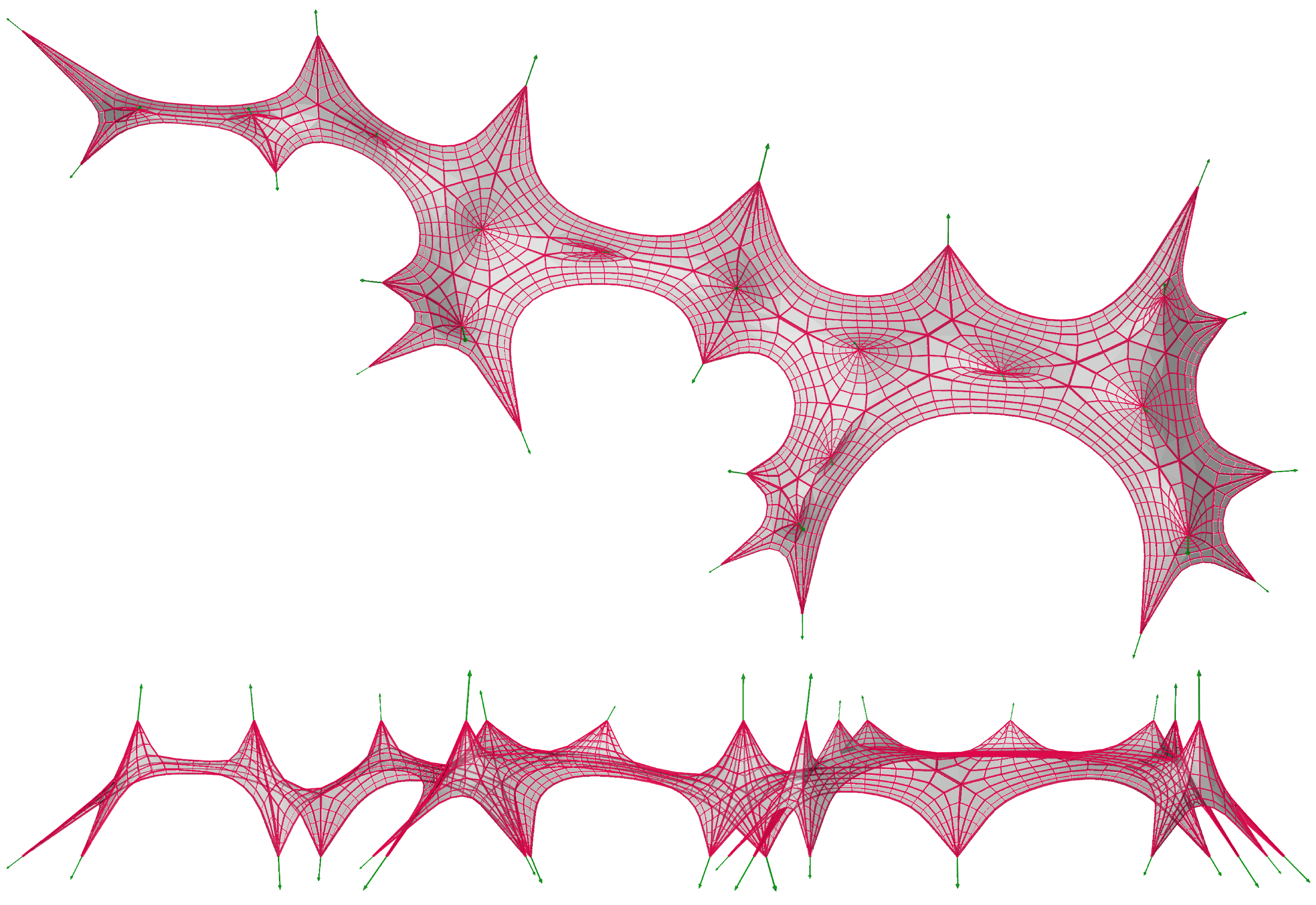}
    \caption{A cable net roof. The cables of this inverse form-found, tension-only structure have a target length of $0.15$ m.}
    \label{fig:rhoen}
    \vskip -0.2in
\end{figure}

We design a self-stressed cable net inspired by the roof of the Rhön Klinikum~\cite{oval_topology_2019}.
Building cable nets from standardized components is important for fabrication efficiency.
Therefore, we calculate a tension-only shape for the net that has a target edge length of $0.15$ m.
Fig.~\ref{fig:rhoen} displays the solution to the inverse form-finding problem.
The goal function $g(f(\theta,G))$ in this problem is:
\begin{minted}{python}
def goal_fn(eq_state):
    diff = (eq_state.lengths - 0.15)**2
    std = jnp.std(eq_state.lengths)
    return jnp.sum(diff) + 0.01 * std
\end{minted}
After modeling the connectivity of the cable net as a \texttt{structure}, we can reuse the code blocks presented in Section~\ref{sec:results:shell} to obtain $\mathbf{q}^{\star}$.
The only requirement is to swap \texttt{goal\_fn} in the body of the loss function \texttt{loss\_fn}.
The composition and interchangeability of such atomic goal functions simplify the formulation of custom inverse form-finding problems with JAX FDM.
We applied a similar approach to generate the planar cable net shown in Fig.~\ref{fig:jax_logo}.
\section{Conclusion}

We presented JAX FDM, an open-source solver that streamlines the design of 3D structures in static equilibrium conditioned on target properties.
Future work will include our solver as a differentiable layer in neural networks to build accurate surrogate models that further accelerate inverse form-finding.

\section*{Acknowledgements}

This work has been supported by the U.S. National Science Foundation under grant OAC-2118201, and by the Institute for Data-Driven Dynamical Design (ID4).


\bibliography{bibliography}
\bibliographystyle{icml2023}




\end{document}